\documentclass[prb,twocolumn,showpacs,preprintnumbers,amsmath,amssymb,superscriptaddress, bibnotes]{revtex4}

\usepackage{color}
\usepackage{graphicx}
\usepackage{dcolumn}
\usepackage{bm}
\usepackage{mathrsfs}
\begin{document}

\title{$^{31}$P and $^{75}$As NMR evidence for a residual density of states at zero energy \\in superconducting BaFe$_2$(As$_{0.67}$P$_{0.33}$)$_2$}

\author{Yusuke Nakai}
\email{nakai@scphys.kyoto-u.ac.jp}
\author{Tetsuya Iye}
\author{Shunsaku Kitagawa}
\author{Kenji Ishida}
\affiliation{Department of Physics, Graduate School of Science, Kyoto University, Kyoto 606-8502, Japan,}
\affiliation{TRIP, JST, Sanban-cho Building, 5, Sanban-cho, Chiyoda, Tokyo 102-0075, Japan,}

\author{\\Shigeru Kasahara}
\affiliation{Research Center for Low Temperature and Materials Sciences, Kyoto University, Kyoto 606-8502, Japan}
\author{Takasada Shibauchi}
\author{Yuji Matsuda}
\affiliation{Department of Physics, Graduate School of Science, Kyoto University, Kyoto 606-8502, Japan,}
\author{Takahito Terashima}
\affiliation{Research Center for Low Temperature and Materials Sciences, Kyoto University, Kyoto 606-8502, Japan}

\date{\today}

\begin{abstract}
$^{31}$P and $^{75}$As NMR measurements were performed in superconducting BaFe$_2$(As$_{0.67}$P$_{0.33}$)$_2$ with $T_c = 30$ K.
The nuclear-spin-lattice relaxation rate $T_1^{-1}$ and the Knight shift in the normal state indicate the development of antiferromagnetic fluctuations, and $T_1^{-1}$ in the superconducting (SC) state decreases without a coherence peak just below $T_c$, as observed in (Ba$_{1-x}$K$_{x}$)Fe$_2$As$_2$. In contrast to other iron arsenide superconductors, the $T_1^{-1} \propto T$ behavior is observed below 4K, indicating the presence of a residual density of states at zero energy. Our results suggest that strikingly different SC gaps appear in BaFe$_2$(As$_{1-x}$P$_x$)$_2$ despite a comparable $T_c$ value, an analogous phase diagram, and similar Fermi surfaces to (Ba$_{1-x}$K$_{x}$)Fe$_2$As$_2$.  
\end{abstract}

\maketitle

The symmetry of the superconducting (SC) order parameter is one of the most important issues in iron pnictide superconductors.~\cite{IshidaJPSJReview} NMR studies on $R$FeAsO (``1111", $R$ is a rare-earth element)~\cite{NakaiJPSJ2008,NakaiDopingDep,Grafe,MukudaJPSJ08,TerasakiFeNMR,MatanoEPL2008,KawasakiPRB2008,KobayashiJPSJ2009} and $A$Fe$_2$As$_2$ superconductors (``122", $A$ is an alkaline-earth element )~\cite{Fukazawa(BaK)122JPSJ2009,Matano(BaK)122,Yashima(BaK)122,JulienEPL,NingJPSJBa(FeCo)2As2DopingDep} have consistently found the absence of a coherence peak, but $T$-dependences of the spin-lattice relaxation rate $T_1^{-1}$ below $T_c$ remain controversial. Although these various behaviors of $T_1^{-1}$ below $T_c$ may reflect that $T_1^{-1}$ in iron pnictides is sensitive to doping levels and/or sample quality, these NMR results~\cite{NakaiJPSJ2008,NakaiDopingDep,Grafe,MukudaJPSJ08,TerasakiFeNMR,MatanoEPL2008,KawasakiPRB2008,KobayashiJPSJ2009,Fukazawa(BaK)122JPSJ2009,Matano(BaK)122,Yashima(BaK)122} show no evidence of a residual density of states (DOS) at zero energy in the SC state.~\cite{KitagawaSr122SCnote} This is in sharp contrast to the case of high-$T_c$ cuprates, where dilute impurities and crystalline defects easily induce a residual DOS at the Fermi level. 

Recently, the penetration depth study on the iron phosphide LaFePO ($T_c\sim$ 6 K) suggested nodal superconductivity,~\cite{FletcherPenetrationLaFePO} which is inconsistent with the prevailing theory of fully-gapped $s_{\pm}$ states.~\cite{MazinPRL2008,KurokiPRL2008,CvetkocicEPL} This different SC state is shown to be explained by changes in pnictogen height acting as a switch between high-$T_c$ nodeless and low-$T_c$ nodal pairing.~\cite{KurokiPRB2009} In this regard, the isovalent P-substituted BaFe$_2$(As$_{1-x}$P$_x$)$_2$ where P substitution is expected to reduce the pnictogen height is interesting in order to investigate the nature of the SC gap function. Isovalent P substitution for As can tune the ground state of BaFe$_2$As$_2$ from antiferromagnetism to superconductivity with a comparable maximum $T_c$ value of 30 K, similar Fermi surfaces,  and a similar phase diagram to (Ba$_{1-x}$K$_x$)Fe$_2$As$_2$ (see Fig.~1).~\cite{JiangBaFe2(AsP)2,KasaharaBaFe2(AsP)2} It is believed that P substitution in BaFe$_2$(As$_{1-x}$P$_x$)$_2$ does not introduce charge carriers, unlike the hole-doped (Ba$_{1-x}$K$_x$)Fe$_2$As$_2$ and electron-doped Ba(Fe$_{1-x}$Co$_x$)$_2$As$_2$. Here, we present NMR evidence for a residual DOS at zero energy in BaFe$_2$(As$_{0.67}$P$_{0.33}$)$_2$, which is fundamentally different from other iron arsenide superconductors. 
Our findings suggest that different SC pairing states are realized in iron pnictides despite comparable high-$T_c$ values, analogous phase diagrams, and similar Fermi surfaces. 

%************************Fig.1***************************************
\begin{figure}[tb]
\begin{center}
\includegraphics[width=6cm]{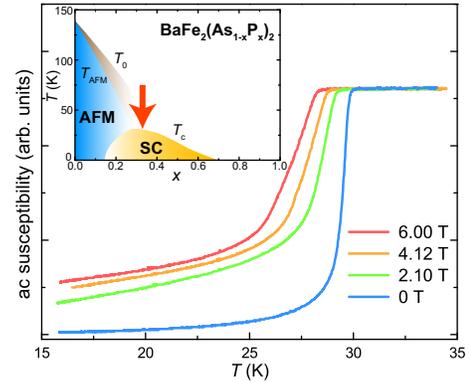}
\end{center}
\caption{(Color online) ac susceptibility of BaFe$_2$(As$_{0.67}$P$_{0.33}$)$_2$ obtained with an NMR coil at $\sim17$ MHz, indicating $T_c(H=0)=30$ K. Inset: schematic phase diagram of BaFe$_2$(As$_{1-x}$P$_x$)$_2$. $T_0$ ($T_{\rm AFM}$) represents the structural (AFM) ordering temperature. The arrow denotes the P substitution $x$ of the present sample with maximum $T_c$.}
\end{figure}
%********************************************************************
Polycrystalline samples of BaFe$_2$(As$_{0.67}$P$_{0.33}$)$_2$, consisting of single crystals with average dimensions of 100 $\times$ 100 $\times$ 50 $\mu$m$^3$, were prepared from mixtures of FeAs, Fe, P, and Ba.~\cite{KasaharaBaFe2(AsP)2} 
The P substitution $x$ was determined by an energy dispersive x-ray. 
ac-susceptibility measurements with an NMR coil show that $T_c(H=0)$ is 30.0 K with a narrow transition width, indicating an excellent sample quality  (see Fig.~1). 

%************************Fig.2***************************************
\begin{figure}[tb]
\begin{center}
\includegraphics[width=6.3cm]{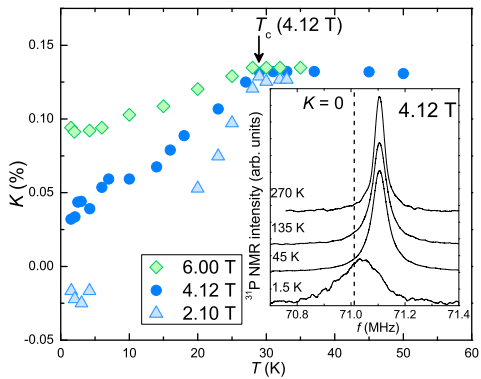}
\end{center}
\caption{(Color online) $^{31}$P Knight shift $K$ for different magnetic fields. Inset shows $^{31}$P NMR spectra in 4.12 T. The spectra have been  offset vertically.}
\end{figure}
%***********************************************************************
The inset in Fig.~2 shows $^{31}$P NMR spectra obtained by sweeping frequency at a fixed magnetic field (4.12 T). 
The Knight shift $K$ was determined with respect to a resonance line in H$_3$PO$_4$ ($K=0$ in Fig.~2). 
$K$ is $T$-independent ($K$ = 0.131 $\pm$ 0.001\%) in the normal state and decreases in the SC state. 
From the magnetic field dependence of $K$, we find that this decrease is dominated by the SC diamagnetic shift in the mixed state. 
The large SC diamagnetic shift prevents us from determining variations in the spin susceptibility in the SC state. 
The $^{31}$P NMR spectra show no evidence for antiferromagnetic (AFM) order at any temperature.~\cite{KitagawaSr122SCnote} 

%************************Fig.3***************************************
\begin{figure}[tb]
\begin{center}
\includegraphics[width=7.2cm]{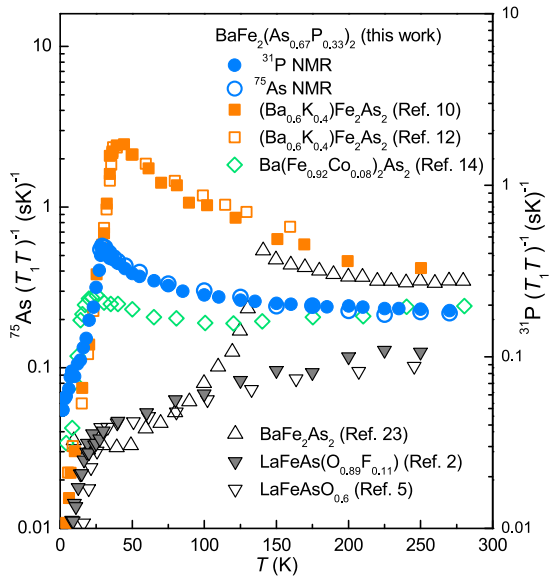}
\end{center}
\caption{(Color online) $^{31}$P (right axis) and $^{75}$As (left axis) $(T_1T)^{-1}$ for BaFe$_2$(As$_{0.67}$P$_{0.33}$)$_2$, along with $^{75}$As $(T_1T)^{-1}$ in optimally doped iron arsenide superconductors.~\cite{NakaiJPSJ2008,MukudaJPSJ08,Fukazawa(BaK)122JPSJ2009,Yashima(BaK)122,NingJPSJBa(FeCo)2As2DopingDep,KitagawaBaFe2As2} }
\label{NormalT1}
\end{figure}
\begin{figure}[tb]
\begin{center}
\includegraphics[width=6.4cm]{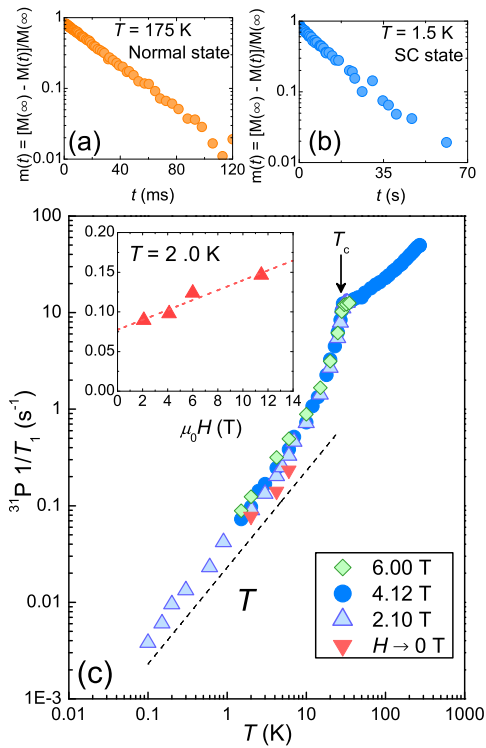}
\end{center}
\caption{(Color online) (a) and (b) show $^{31}$P nuclear-spin relaxation curves for 175 K and 1.5 K in 4.12 T, exhibiting clear single exponential behavior. 
(c) $^{31}$P $T_1^{-1}$ under 2.1, 4.12, and 6.0 T.  Inset in panel (c) shows $^{31}$P $T_1^{-1}$ vs magnetic field at 2 K; a linear fit is used to extrapolate $T_1^{-1}$ for $H \to$ 0 T.}
\label{SCT1}
\end{figure}

%***********************************************************************
The $^{31}$P ($^{75}$As) nuclear spin-lattice relaxation rate $T_1^{-1}$ was determined by fitting the time dependence of the spin-echo intensity to a theoretical $I = 1/2$ (3/2) curve after a saturation pulse. 
Good fits are obtained throughout the temperature range 100 mK - 270 K (see Fig.~\ref{SCT1}), 
indicating a homogeneous electronic state in BaFe$_2$(As$_{0.67}$P$_{0.33}$)$_2$. 
In contrast to the $T$-independent Knight shift, $(T_1T)^{-1}$ for $^{31}$P and $^{75}$As increases significantly on cooling down to $T_c$ (see Fig.~\ref{NormalT1}). 
Since $(T_1T)^{-1}$ is related to the sum of $\bm{q}$-averaged dynamical spin susceptibility $\chi(\bm{q})$ at low energy and the Knight shift is proportional to $\chi(\bm{q}=0)$, this $(T_1T)^{-1}$ result clearly shows the development of low-lying spin fluctuations at finite wave-vectors $\bm{Q}$ away from zero. 

Next, we compare spin dynamics in BaFe$_2$(As$_{0.67}$P$_{0.33}$)$_2$ with those in (Ba$_{0.6}$K$_{0.4}$)Fe$_2$As$_2$, Ba(Fe$_{0.92}$Co$_{0.08}$)$_2$As$_2$, LaFeAs(O$_{0.89}$F$_{0.11}$), and LaFeAsO$_{0.6}$ superconductors, which are all near optimum doping. 
As in BaFe$_2$(As$_{0.67}$P$_{0.33}$)$_2$, the development of low-energy AFM fluctuations is inferred from an increase in $(T_1T)^{-1}$ in (Ba$_{1-x}$K$_x$)Fe$_2$As$_2$ (Refs. 10-13)~\cite{Fukazawa(BaK)122JPSJ2009,Yashima(BaK)122,Matano(BaK)122,JulienEPL} and Ba(Fe$_{1-x}$Co$_x$)$_2$As$_2$,~\cite{NingJPSJBa(FeCo)2As2DopingDep,LaplaceCo-dopedBaFe2As2} suggesting the importance of AFM fluctuations for superconductivity (see Fig.~3). 
In contrast, low-lying AFM fluctuations are suppressed above $T_c$ in optimally doped LaFeAs(O$_{0.89}$F$_{0.11}$) (Refs. 2 and 3)~\cite{NakaiJPSJ2008,NakaiDopingDep} and LaFeAsO$_{0.6}$,~\cite{MukudaJPSJ08,TerasakiFeNMR} implying different nature of low-lying AFM fluctuations in 122 and 1111 superconductors. 
Interestingly, the $T$-dependences of $(T_1T)^{-1}$ differ even among the Ba122 superconductors; $(T_1T)^{-1}$ in (Ba$_{0.6}$K$_{0.4}$)Fe$_2$As$_2$ exhibits the most significant $T$-dependence, there is only a slight enhancement below about 50 K in Ba(Fe$_{0.92}$Co$_{0.08}$)$_2$As$_2$, and an intermediate $T$-dependence of $(T_1T)^{-1}$ is observed in BaFe$_2$(As$_{0.67}$P$_{0.33}$)$_2$. 
These variations indicate that low-lying AFM fluctuations are sensitive to composition, and this is attributable to different nesting conditions in their Fermi surfaces.~\cite{KasaharaBaFe2(AsP)2,HashimotoPenetrationBaFe2(AsP)2} 
Although these materials have similar Fermi surfaces with three hole sheets and two electron sheets, band calculations suggest that P and K substitutions would modify their Fermi surfaces in different ways~\cite{KasaharaBaFe2(AsP)2,HashimotoPenetrationBaFe2(AsP)2}; P substitution makes the hole sheets having stronger $c$-axis warping, and thus nesting is only possible in restricted regions along the $c$ direction, while K-doping makes the hole sheets larger and the electron sheets smaller, resulting in only one electron- and hole-sheet among five contributing to the nesting. 

Figure~3 highlights systematic variations in $(T_1T)^{-1}$ from hole- to electron-doped arsenides. The large values of $(T_1T)^{-1}$ in the hole-doped (Ba$_{0.6}$K$_{0.4}$)Fe$_2$As$_2$ at high temperatures most likely originate from the enhanced DOS at the Fermi level inferred from band calculations.~\cite{HashimotoPenetrationBaFe2(AsP)2} 
In contrast, in the electron-doped Ba(Fe$_{0.92}$Co$_{0.08}$)$_2$As$_2$, LaFeAs(O$_{0.89}$F$_{0.11}$), and LaFeAsO$_{0.6}$, $(T_1T)^{-1}$ is smaller and decreases on cooling at high temperatures. 
This is attributable to a hole Fermi surface with a high DOS just below the Fermi level.~\cite{IkedaJPSJ2008} 
Thus, the variations in $T$-dependences of $(T_1T)^{-1}$ suggest that spin dynamics probed by NMR are strongly related to the nature of the Fermi surfaces. 

%************************Fig.4***************************************
%********************************************************************
%************************Fig.5***************************************
\begin{figure}[tb]
\begin{center}
\includegraphics[width=6cm]{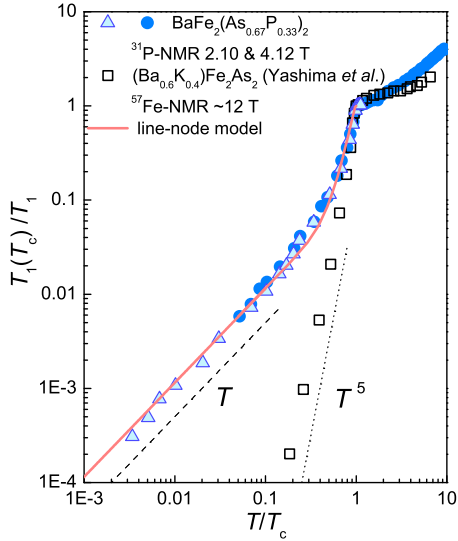}
\end{center}
\caption{(Color online) Normalized $^{31}$P $T_1^{-1}$ for BaFe$_2$(As$_{0.67}$P$_{0.33}$)$_2$ and (Ba$_{0.6}$K$_{0.4}$)Fe$_2$As$_2$.~\cite{Yashima(BaK)122} The solid curve shows a calculation for a line-node gap function $\Delta(\phi)=\Delta_0\sin{(2\phi)}$ with $2\Delta_0 = 6k_BT_c$, $N_{\rm res}/N_0=0.34$.}
\label{SCT1Compare}
\end{figure}
%********************************************************************
In the SC state of BaFe$_2$(As$_{0.67}$P$_{0.33}$)$_2$, we observed a marked $T_1^{-1}$ behavior distinctly different from other iron arsenides. 
Figure~\ref{SCT1}~(c) shows that $^{31}$P $T_1^{-1}$ decreases sharply just below $T_c$ and there is no evidence for a coherence peak. 
We find $T_1^{-1}\propto T$ below $\sim4$ K down to 100 mK, showing the existence of a residual DOS at zero energy. 
Below about 10 K ($\approx0.3T_c$), $T_1^{-1}$ depends on the magnetic field [see Fig.~4~(c)]. By extrapolating to zero field, we still find $T_1T=$ const behavior below $\sim4$ K, excluding the possibility that the $T_1T=$ constant behavior is attributable to applied fields. 
The existence of a residual DOS in $H\to$ 0 inferred from NMR is consistent with thermal conductivity results on BaFe$_2$(As$_{0.67}$P$_{0.33}$)$_2$ that reveal a finite thermal conductivity divided by temperature $\kappa_0/T$ at $T\to0$ under zero field.~\cite{HashimotoPenetrationBaFe2(AsP)2} 
Our findings may also be related to the recent specific heat results on Ba(Fe$_{1-x}$Co$_x$)$_2$As$_2$.~\cite{MuSpecificHeatBaFeCo122}

There are two possible contributions to the field dependence of $T_1^{-1}$ arising from quasiparticle excitations in the mixed state: localized quasiparticles inside vortex cores, and delocalized quasiparticles originating from nodes in the SC gap. 
Since the applied fields are much smaller than $H_{\rm c2}\approx52$ T,~\cite{HashimotoPenetrationBaFe2(AsP)2} 
localized quasiparticles inside vortex cores are unlikely to contribute to the relaxation at these fields ($H/H_{c2}\leq0.22$). 
In fact, the thermal conductivity measurements also show a sizable field dependence of  $\kappa_0(H)/T$ at low fields, attaining nearly 70\% of the normal state value even at 0.2 $H_{\rm c2}$, indicative of the predominance of delocalized quasiparticles.~\cite{HashimotoPenetrationBaFe2(AsP)2} 
These results suggest that the field dependence of $T_1^{-1}$ is attributable to the contribution from delocalized quasiparticles originating from nodes in the gap, compatible with the existence of a residual DOS in the SC state. 

The existence of a residual DOS in the SC state of BaFe$_2$(As$_{0.67}$P$_{0.33}$)$_2$ differs strikingly from other iron arsenides, including 1111~\cite{NakaiJPSJ2008,NakaiDopingDep,Grafe,MukudaJPSJ08,TerasakiFeNMR,MatanoEPL2008,KawasakiPRB2008,KobayashiJPSJ2009} and 122 superconductors,~\cite{Fukazawa(BaK)122JPSJ2009,Matano(BaK)122,Yashima(BaK)122,FukazawaK122} instead resembling the cuprates and heavy fermion superconductors, which have nodes in their SC gap. 
As shown in Fig.~5, $^{57}$Fe $T_1^{-1}$ in (Ba$_{0.6}$K$_{0.4}$)Fe$_2$As$_2$ near optimal doping exhibits a $T^5$-like dependence well below $T_c$.~\cite{Yashima(BaK)122} The fairly large exponent of $\sim5$ in $T_1^{-1}$ implies a fully gapped state in (Ba$_{0.6}$K$_{0.4}$)Fe$_2$As$_2$, and was analyzed in terms of $s_{\pm}$-wave superconductivity with two gaps.~\cite{Yashima(BaK)122} 
In the $s_{\pm}$ models with unitary scatterings,~\cite{ParkerPRB2008,BangPRB2008} in-gap states at zero energy can be induced without severely reducing $T_c$ and can in principle account for the $T_1^{-1}\propto T$ behavior.  
However, the penetration depth $\Delta\lambda(T)\propto T$ at low temperatures observed in BaFe$_2$(As$_{0.67}$P$_{0.33}$)$_2$ is incompatible with these models,~\cite{VorontsovPRB2009,BangEPL2009} but is consistent with line nodes in the gap.~\cite{HashimotoPenetrationBaFe2(AsP)2}
In a symmetry-imposed nodal superconductor such as described by a line-node model $\Delta=\Delta_0\sin{(2\phi)}$, $T_1^{-1}\propto T^3$ in the clean limit, but impurity scattering induced near gap nodes gives rise to an enhanced $T_1^{-1}$ that is proportional to temperature at low temperatures.~\cite{IshidaJPSJ1993} P dopants in BaFe$_2$(As$_{0.67}$P$_{0.33}$)$_2$ can provide a source for nonmagnetic scatterers, consistent with the existence of Fermi-level DOS. 
In fact, the $T_1^{-1}$ data of BaFe$_2$(As$_{0.67}$P$_{0.33}$)$_2$ are well fit by the line-node model with $2\Delta_0/k_BT_c = 6$ and $N_{\rm res}/N_0 = 0.34$, where $N_{\rm res}/N_0$ is the ratio of a residual DOS in the SC state to that in the normal state (see Fig.~5). 
Therefore, NMR, penetration-depth, and thermal conductivity results on BaFe$_2$(As$_{0.67}$P$_{0.33}$)$_2$ are consistent with the existence of line nodes in the SC gap. 

In the case of nodes imposed by symmetry, a residual DOS at zero energy is a natural consequence of the impurity scattering that broadens the nodes. However, this is not clear for the case of ``accidental" nodes not imposed by symmetry, as in the extended $s$-wave case because impurity scattering will lift the nodes and produce a more isotropic gap.~\cite{MishraPRB2009} Our experimental findings impose significant constraints on candidate theoretical descriptions of iron pnictide superconductivity. 

The strikingly different behavior of $T_1^{-1}$ in the SC state between BaFe$_2$(As$_{0.67}$P$_{0.33}$)$_2$ and (Ba$_{0.6}$K$_{0.4}$)Fe$_2$As$_2$ is surprising because they have comparable $T_c$ values and similar Fermi surfaces. 
This implies that high-$T_c$ superconductivities with and without nodes are nearly degenerate and that subtle differences can significantly modify the nature of superconductivity.~\cite{KurokiPRL2008,KurokiPRB2009,MishraPRB2009,SeoPRL2008,Chubukov2009} 
Changes in the pnictogen height may lead to nodal superconductivity, but its $T_c$ should be much lower than that of fully-gapped $s_{\pm}$ states.~\cite{KurokiPRB2009} However, this is not the case in BaFe$_2$(As$_{0.67}$P$_{0.33}$)$_2$. 
One notable difference inferred from band calculations~\cite{KasaharaBaFe2(AsP)2,HashimotoPenetrationBaFe2(AsP)2} is that BaFe$_2$(As$_{0.67}$P$_{0.33}$)$_2$ may be more three-dimensional than (Ba$_{0.6}$K$_{0.4}$)Fe$_2$As$_2$, suggesting that a three-dimensional model is required. 
Reconciling their comparable $T_c$ values and fundamentally different SC gap functions needs further theoretical and experimental works. 
To clarify this problem, identifying the nodes' location on Fermi surfaces will be crucial. 

In conclusion, our NMR results show the existence of a residual DOS at zero energy in BaFe$_2$(As$_{0.67}$P$_{0.33}$)$_2$. 
This is consistent with the presence of nodes in the SC gap inferred from a field dependence of $T_1^{-1}$, and from thermal conductivity and penetration depth results. 
Our results suggest that strikingly different SC gaps appear in iron pnictides despite comparable high-$T_c$ values, analogous phase diagrams, and similar Fermi surfaces. 

We thank K. Kitagawa, D.~C.~Peets, S. Yonezawa, H.~Ikeda and Y. Maeno for experimental support and discussions. 
This work was supported by the Grants-in-Aid for Scientific Research on Innovative Areas ``Heavy Electrons" (No. 20102006) from MEXT, for the GCOE Program ``The Next Generation of Physics, Spun from Universality and Emergence" from MEXT, and for Scientific Research from JSPS. Y.N. is supported by JSPS.

%\bibliography{D:/thesis/LaOFeX/1111,D:/thesis/LaOFeX/122,D:/thesis/LaOFeX/theory,D:/thesis/LaOFeX/NMR,D:/thesis/LaOFeX/LaFePO}

\end{document}